\documentclass[twocolumn]{emulateapj}
\usepackage{amsmath,bm}
\usepackage{amssymb}
\usepackage{epsfig}
\usepackage{color}
\begin{document}

\title{Gamma-ray production in the extended halo of the Galaxy and possible implications for the origin of Galactic cosmic rays}

\author{Ruo-Yu Liu\altaffilmark{1}, Huirong Yan
\altaffilmark{1,2}, Xiang-Yu Wang\altaffilmark{3}, Shi Shao\altaffilmark{4}, Hui Li\altaffilmark{5}}
\altaffiltext{1}{Deutsches Elektronen Synchrotron (DESY), Platanenallee 6, D-15738 Zeuthen, Germany;ruoyu.liu@desy.de; huirong.yan@desy.de}
\altaffiltext{2}{Institut f\"ur Physik und Astronomie, Universit\"at Potsdam, D-14476 Potsdam, Germany}
\altaffiltext{3}{School of Astronomy and Space Science, Nanjing University, Nanjing
210093, China}
\altaffiltext{4}{Institute for Computational Cosmology, Department of Physics, Durham University, South Road Durham DH1 3LE, UK}
\altaffiltext{5}{Department of Physics, Kavli Institute for Astrophysics and Space Research, Massachusetts Institute of Technology, Cambridge, MA 02139, USA}

\begin{abstract}
Various studies have implied the existence of a gaseous halo around the Galaxy extending out to $\sim 100\,$kpc. Galactic cosmic rays (CRs) that propagate to the halo, either by diffusion or by convection with the possibly existing large-scale Galactic wind, can interact with the gas therein and produce gamma-rays via proton-proton collision.
{We calculate the cosmic ray distribution in the halo and the gamma-ray flux, and explore the dependence of the result on model parameters such as diffusion coefficient, CR luminosity, CR spectral index.} We find that the current measurement of isotropic gamma-ray background at $\lesssim$TeV with  Fermi Large Area Telescope already approaches a level that can provide interesting constraints on the properties of Galactic cosmic ray {(e.g., with CR luminosity $L_{\rm CR}\leq 10^{41}\,$erg/s)}. We also discuss the possibilities of the Fermi bubble and IceCube neutrinos originating from the proton-proton collision between cosmic rays and gas in the halo, as well as the implication of our results for the baryon budget of the hot circumgalactic medium of our Galaxy. Given that the isotropic gamma-ray background is likely to be dominated by unresolved extragalactic sources, future telescopes may extract more individual sources from the IGRB, and hence put even more stringent restriction on the relevant quantities (such as Galactic cosmic ray luminosity and baryon budget in the halo) in the presence of a turbulent halo that we consider.
\end{abstract}

\maketitle

\section{Introduction}
The question that how much cosmic-ray (CR) luminosity of our Galaxy is needed to maintain the observed CR flux was firstly raised by \citep{Ginzburg64}. {The answer to this question not only provides a clue to the origin of Galactic CRs from the point of view of energetics that the sources can provide, but may also reflect the propagation nature of CRs inside the Galaxy and hence is also of great interest to other relevant fields such as the interstellar medium, plasma astrophysics and so on}. Based on the measurement of the local CR energy density and the so-called ``grammage'' {(i.e., the amount of matter traversed by GCRs before reaching the Earth) which represents the propagation time of CRs}, the CR luminosity of our Galaxy above 1~GeV is usually found to be $3\times 10^{40}\,{\rm erg/s}<L_{\rm CR}<3\times 10^{41}\,{\rm erg/s}$ in the framework of the leaky-box model \citep{Ginzburg64, Drury89, Berezinsky90, Dogiel02, Strong10, Drury17}, where the uncertainty is due to the statistical error in measurement and the selection of the CR propagation model. The estimated value of the CR luminosity provides an important clue to the species of the CR accelerators and their efficiency.

CRs can be studied through gamma-rays produced in the hadronuclear interaction or proton--proton collision (hereafter $pp$ collision) between CRs and {diffuse baryonic material}. This method has been widely used among the community, for example, to derive the CR distribution in the Galactic plane \citep[e.g.][]{Fermi09_DGE} or as indicators of CR accelerators \citep[e.g.][]{Fermi13_SNR, HESS_GC16}. {Interestingly,  recent observations of ion absorption lines against background quasars \citep[e.g.][]{Nicastro02, Rasmussen03, Miller13, Fang15, ZhengY17} and emission lines \citep[e.g.][]{Henley12, Henley13, Miller15} at high Galactic latitudes along different line-of-sights {suggest the existence of a hot baryon gas halo surrounding the Galaxy, which is also known as the circumgalactic medium (CGM)}. The existence of the CGM is also supported by various indirect observations \citep[e.g.][]{Stanimirovic02,Fox05,Grcevich09, Putman11}.} {The total mass of the CGM is inferred to be several times $10^{10}\,M_\odot$ within the virial radius (250~kpc) of the Galaxy. It} serves as a target of baryons and can interact with CRs that escape our Galaxy at past via the $pp$ collision and give rise to gamma-ray photons. The gamma-ray photons {produced} at high Galactic latitudes {may/will contribute to the isotropic gamma-ray background (IGRB) as measured} by various instruments such as the \emph{SAS-2} satellite \citep{Fichtel75, Fichtel78}, EGRET on board the \emph{Compton Observatory} \citep{Sreekumar98, Strong04}, and the Fermi Large Area Telescope (\emph{Fermi}-LAT, \citealt{Fermi10_IGRB}) up to 100\,GeV.

The $pp$ collision between the gas and cosmic rays including the consequent production of gamma rays in the Galactic halo has been considered by various authors \citep{Stecker77, DP00, Feldmann13, Ahlers14, Taylor14}. {Recently, {\it Fermi}-LAT updated the spectrum measurement on IGRB up to to 820\,GeV \citep{Fermi15}. Extrapolating the prediction of earlier works to this sub-TeV energy range and comparing them to the new \emph{Fermi}-LAT measurement allows to rule out some of the physical setups/parameters ranges considered by the authors. More recently, \citet{Kalashev16} addressed the sub-TeV measurement of the extragalactic gamma-ray background (EGB) which also includes the extragalactic point-source contribution in addition to the IGRB. Although their result is consistent with the sub-TeV EGB flux, it is likely to be in tension with the sub-TeV IGRB flux.}

In this work, we will study gamma rays produced in the interactions between the CRs and the gas in the halo, and use the results to put constraints on model parameters such as the CR luminosity of the Galaxy. We take into account diffusion in turbulent halo and the convection by the large-scale Galactic wind that might exist and the subsequent adiabatic cooling of CRs in the wind, which were neglected in some previous studies. {We will obtain constraints, which are independent of direct measurements on CRs, on various properties of Galactic CR as well as the diffusion coefficient in the halo and the mass of CGM} from the IGRB flux at $\lesssim$TeV energy measured by {\it Fermi}-LAT.

This paper is organized as follows. We calculate the cosmic ray distribution and the gamma-ray production in the halo in Section 2. In Section 3, the results are presented and the constraint on the Galactic CR luminosity is discussed. We discuss the anisotropy of the gamma-ray intensity, the neutrino emission and the baryon budget of the Galaxy in Section 4 and summarize the work in Section 5.

\section{Cosmic ray Propagation and Gamma-ray Production}

\begin{figure*}[htbp]
\centering
\includegraphics[width=0.9\textwidth]{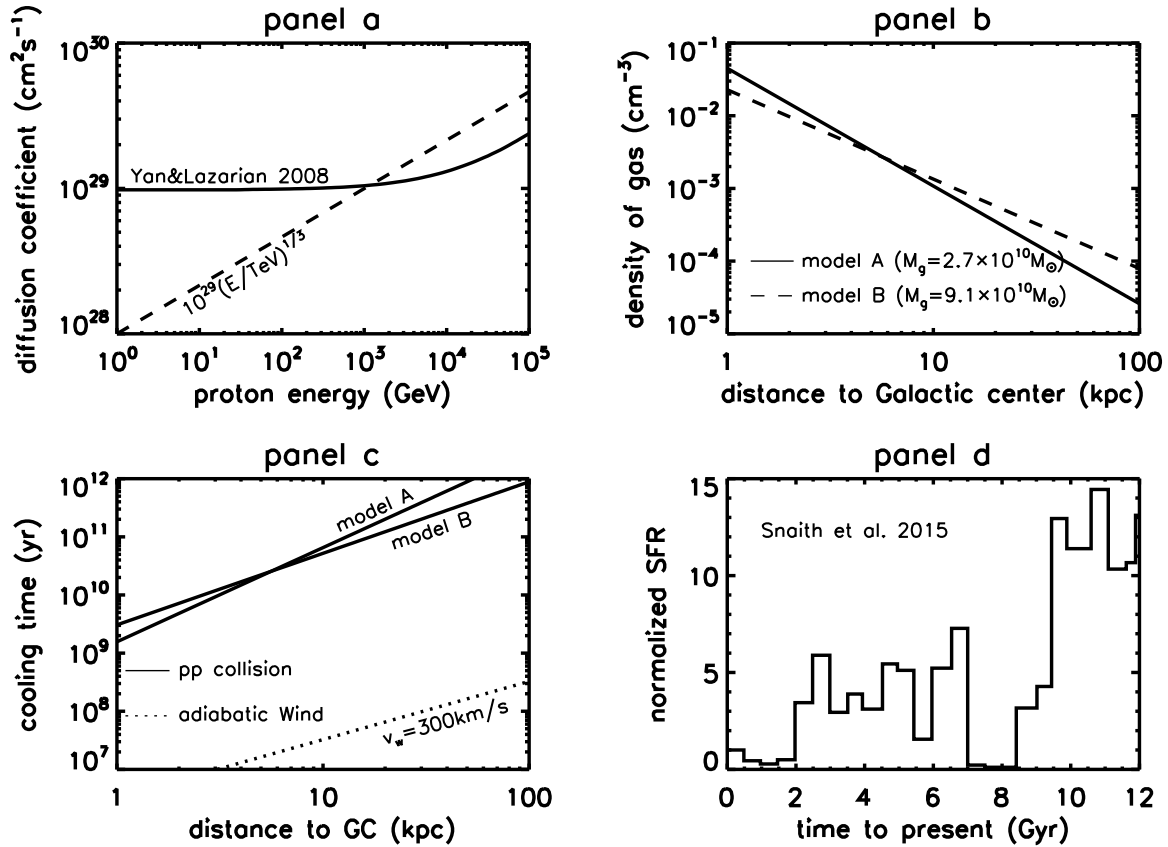}
\caption{{Model parameters and basic properties: {\bf top left (panel~a):} diffusion coefficient in the halo obtained by \citep{Yan08}, the usually adopted form of diffusion coefficient in Galactic disk $D(E)=10^{29}(E/\rm \,TeV)^{1/3}\rm\,cm^2s^{-1}$ is also plotted for reference; {\bf top right (panel~b):} density profile of hot gas in the halo; {\bf bottom left (panel~c):} cooling time scale due to $pp$ collision (solid lines) and adiabatic loss due to the expansion of the Galactic wind (dotted line); {\bf bottom right (panel~d):} SFH obtained by \citep{Snaith15}, normalized to 1 at present day.}}\label{fig:setup}
\end{figure*}

{To calculate the gamma-ray flux originating from the halo, we need the distribution of CRs and gas in the halo. In this section, we first study the propagation and evolution of CRs in the halo. In our model we consider CR propagation by diffusion, convection by the Galactic wind and the cooling of the particles. Then, we use the CR and the gas distribution in the CGM to calculate the gamma-ray emissivity in the halo and the total gamma-ray flux at the Sun's position in the Galaxy, for comparison with observations.}

The transport of CRs injected from a point source is regulated by the equation \citep{Berezinsky06b}
\begin{equation}\label{eq:transport}
\frac{\partial n}{\partial t}+ \bm{v}_w\nabla_{\bm{r}} n -D(E)\nabla_{\bm{r}}^2n-\frac{\partial}{\partial E}\left[b(E,t)n \right]=Q(E,t)\delta^3(\bm{r}-\bm{r}_g)
\end{equation}
where $n(\bm{r},E,t)$ is the differential density of CRs at time $t$ and at $\bm{r}=(x,y,z)$ which is the Cartesian coordinates of a certain position in space measured in the comoving frame. $\bm{r}_g$ is the coordinates of the point source. We define the coordinates of the Galactic center and the Earth to be $\bm{r}_C=(0,0,0)$ and $\bm{r}_E=(8.5\,{\rm kpc},0,0)$ respectively, and $t=0$ at the present time.

The second term represents the effect of the convection by a large-scale Galactic wind which could be launched by the pressure gradient of CRs \citep[e.g.][]{Breitschwerdt91, Zirakashvili96,  Hanasz13, Salem14}. {The Galactic wind may be alternatively launched by supernova explosions, as studied in \citep{Dubois08, Sarkar15, Fielding17}}. $\bm{v}_w$ is the velocity of the wind, the value of which is found to increase with the distance {from} the Galactic plane. According to the reference model in \citet{Zirakashvili96} or fiducial model in \citet{Recchia16}, the wind velocity finally approaches an asymptotic value of $\sim 300$km/s at $\sim 100\,$kpc. The wind velocity in other studies is also a few times $100\,$km/s at $\sim 100\,$kpc. { Thus, we take $v_w=300\,$km/s as the fiducial value for the wind velocity in the calculation and assume it constant throughout the halo.} Such a treatment will overestimate the adiabatic cooling of CRs at smaller radius (see discussion in Section 3). 

The third term considers the CR diffusion with the diffusion coefficient $D(E)=cl_{\rm mfp}(E)/3$, where $l_{\rm mfp}$ is the mean free path of CRs. For simplicity, the diffusion coefficient is assumed to be independent on both time and space. The diffusion coefficient in the halo is an important parameter to our calculation but, unfortunately, there is still some uncertainties. {On one hand, the turbulence might be quite weak and hence the diffusion coefficient could be large in the extended halo, if the injection of turbulence into the halo originates from the Galactic plane. On the other hand, CRs may also stream down their pressure gradient by scattering off self-excited Alfv{\'e}n waves \citep[e.g.][]{Kulsrud69, Skilling71}. The self-regulated transport could lead to a small diffusion efficient if the CR density is sufficiently high and the wave damping process is weak. As pointed out in earlier studies \citep{Yan02,Farmer04}, the streaming instability may be suppressed at high energies such as multi-TeV where we mainly concern due to a low CR density at such high energies and also due to the wave damping by the nonlinear Landau damping in the hot plasma. In this work, we do not incorporate the influence of the streaming instability in the calculation, instead of that we follow the mean free path of CRs calculated by \citet{Yan08}, which is based on the current understanding of the Galactic turbulence and calculated with nonlinear theory, to calculate the diffusion coefficient for the entire halo (as is shown in panel~a of Fig.~\ref{fig:setup}). We note that although CR convection by the Galactic wind and diffusion may be somehow related (e.g., the streaming instability of CRs), as a phenomenological study we treat them (i.e., $v_w$ and $D$) as independent parameters in this work}.  We will discuss in detail the dependence of our result on the diffusion coefficient below. 

The last term on the left-hand side describes the effect of continuous energy loss where 
\begin{equation}
\begin{split}
b(E,r,t)&=-\frac{dE}{dt}=\kappa\sigma_{pp}(E)n_g(r)cE+(v_w/r)E\\
&=6\times 10^{-7}\left(\frac{n_g}{10^{-3}\rm cm^{-3}}\right)\left(\frac{E}{10^{12}\rm eV}\right){\rm eV/s}\\
&+10^{-3}\left(\frac{v_w}{300{\rm km/s}}\right)
\left(\frac{r}{10\,\rm kpc}\right)^{-1}\left(\frac{E}{10^{12}\,\rm eV}\right)\rm eV/s
\end{split}
\end{equation}
is the energy loss rate due to $pp$ collision and {the adiabatic expansion of the wind}. $\kappa\simeq 0.5$ is the inelasticity of the $pp$ collision and $\sigma_{pp}$ is the cross section which is about 40\,mb at several TeV { \citep{Kelner06}}. $n_g(r)$ is the gas density profile in the halo and is important to the gamma-ray production since it is proportional to the $pp$ collision rate. \citet{Miller15} analyzed the \ion{O}{7} and \ion{O}{8} emission lines considering different plasma conditions in the halo, and obtained a total mass of the hot gas within the virial radius ranging from $2.7\times 10^{10}M_\odot$ to $9.1\times 10^{10}M_\odot$ under the best-fit parameters. {We then employ the so-called $\beta$-model in the form of $n_g(r)=n_0(r/1\rm\,kpc)^{-3\beta}$, with two sets of parameters which correspond to the lower bound of the total gas mass, i.e., $n_0=0.045\,\rm cm^{-3}$ and $\beta=0.54$ (hereafter, model A and the fiducial case), and the upper bound of  total gas mass, i.e., $n_0=0.023\,\rm cm^{-3}$ and $\beta=0.41$ (hereafter, model B) obtained in their work ({see panel~b in Fig.~\ref{fig:setup}}).} Note that the gas density depends on the metallicity which is assumed to be $Z=0.3Z_\odot$ where $Z_\odot$ is the solar metallicity. This assumption is consistent with the residual pulsar dispersion measure toward the Large Magellanic cloud \citep{Miller15} and simulations of Galactic corona \citep{Toft02, Cen06, Cen12}. Given a density of $10^{-3}-10^{-5}\,\rm cm^{-3}$ in the halo, we expect the energy loss dominated by the adiabatic loss due to the expansion of the Galactic wind. { We plot the CR cooling time, calculated as $E/b(E,t,r)$ for different cooling mechanisms in the panel~c of Fig.~\ref{fig:setup}}.

The term on the right-hand side of { Eq.~\ref{eq:transport}} represents the injection of CRs, which are assumed to consist of pure protons in this work, from a point source located at $\bm{r}_g$. { We assume the injection rate at time $t$ for CR with energy $E$ follows the form}
\begin{equation}
Q(E,t)=S(t)Q_0(E)=S(t)N_0(E/{1\rm \,GeV})^{-p}{\rm exp}(-E/E_{\rm max}),
\end{equation}
where $S(t)$ describes the CR injection history. $p$ is the slope of the injection spectrum. {The locally observed slope of CR spectrum is about 2.7, implying an injection slope of 2.4 given a diffusion coefficient $D(E)\propto E^{1/3}$ in the ISM \citep{Aguilar16}. On the other hand, recent gamma-ray observation of various molecular clouds \citep{Neronov17} and diffuse gamma-ray emission from the inner Galaxy \citep{Yang16} suggest a slope of CR spectrum to be $\sim 2.3-2.5$, implying an injection slope of $2.0-2.2$. Thus, we take $p=2.2$ as a fiducial value and explore the influence of $p$ in the range of $2.0-2.4$. $E_{p,\rm max}$ represents the cutoff energy in the injection spectrum, and we assume $E_{p,\rm max}=10^{15}\,$eV which is comparable to the break in the measured CR spectrum (or the so-called ``knee'').}  {Since CRs are generally believed to be accelerated via strong shocks of supernova remnants, we assume that the CR injection history follow roughly the star formation history (SFH) of the Galaxy. 
\citet{Snaith15} derived the SFH of our Galaxy by fitting their chemical evolution model to a large sample of stellar abundances and we adopt this one for $S(t)$ which is shown in {panel~d of} Fig.~\ref{fig:setup}. Our adopted SFH is also consistent in general with the SFH for Milky Way-sized galaxies derived from abundance matching techniques \citep{Behroozi13}. Given the present CR luminosity to be $L_{\rm CR,0}$ and assume CRs are injected homogeneously from the Galactic plane with a radius of $R_{\rm Gal}=15$~kpc and a negligible thickness (i.e., $\bm{r}_g=(x_g,y_g,0)$ with $\sqrt{x_g^2+y_g^2}<15\,$kpc), we can find $N_0$ from $\pi R_{\rm Gal}^2\int_{1\rm GeV}^{\infty}EQ_0(E)dE=L_{\rm CR,0}$.
We note that the source of CRs from the Galactic plane is probably not homogeneously distributed and the disk size also evolves with redshift. However, since we are concerned with the CR distribution in a 100\,kpc scale halo, the dependence on the distribution of CR sources in the Galactic plane or on the disk size is not significant. {The input model parameters and their values in the fiducial case are summarized in Table.~\ref{tab:parameters}.}

Following the method of \citep{Berezinsky06b} to solve { Eq.~(\ref{eq:transport})} in the Fourier space and transform it back to real space, we obtain the analytical solution of Eq.~(\ref{eq:transport}) for a point source located at $\bm{r}_g$
\begin{equation}\label{eq:crdis}
\begin{split}
n(t,\bm{r},E;\bm{r}_g)&=\frac{\pi^{3/2}}{\left(2\pi\right)^3}\int_{t_g}^tdt'Q(\mathcal{E}',t')\\
&\times \frac{{\rm exp}\left[-(\bm{r}-\bm{r_g}-\bm{s})^2/4\lambda(E,t')\right]}{\lambda(E,t')^{3/2}}\\
&\times {\rm exp}\left[\int_{t'}^tdt''\frac{\partial b(\mathcal{E}'',t)}{\partial \mathcal{E}''} \right]
\end{split}
\end{equation}
where $\bm{s}=\int_{t'}^{t}\bm{v}dt''$ and $\lambda(E,t')=\int_{t'}^{t}D(\mathcal{E}'')dt''$. Here
$\mathcal{E}(t'',E)$ or $\mathcal{E}(t',E)$ means the energy of a CR at time $t''$ or $t'$ which has energy $E$ at the present time. The above solution is the CR density from a point source at $\bm{r}_g$, so we need to further integrate over the Galactic plane in order to obtain the contribution from the whole Galaxy. We set the earliest injection of CRs started at 12~Gyr ago (i.e., $t_g=-12\,$Gyr), which corresponds to a redshift of 4. Since we assume CR injection history follows the SFH of the Galaxy, our selection of $t_g$ includes most CRs injected in the history {(see panel~d of Fig.~\ref{fig:setup})}. Besides, CRs injected at earlier time have propagated to a quite large distance {which is $\sim 2\sqrt{Dt}\simeq 120\,$kpc for diffusion and $\sim v_wt\sim 3.6\,$Mpc for convection given a time $t=12$Gyr}, and hence have small contribution to gamma rays. Thus, considering earlier injection can barely change the results but increase the calculation time. Then, we can obtain the CR density distribution in the halo at present time ($t=0$)
\begin{equation}
N(\bm{r},E)=\int\int n(t=0,\bm{r},E;\bm{r}_g)dx_gdy_g
\end{equation}
In Fig.~\ref{fig:crden}, we show the distribution of 1\,TeV CR energy density along the $z-$direction, i.e., the axis perpendicular to the Galactic plane passing through the Galactic center (GC) at $r_C$ (black curves), the one passing through the Earth at $r_E=$ (blue curves), and the one passing through the edge of the Galactic plane at (15kpc, 0, 0) (red curves). {Compared to the pure diffusion case, the presence of a Galactic wind will transport CRs to farther distances within the same amount of time. The adiabatic cooling of CRs in the expanding wind also leads to an extra energy loss of CRs especially at small Galactocentric radius. As a result, the CR density within 100\,kpc in the case without wind is higher than that in the case with wind.}

\begin{figure}
\includegraphics[width=0.9\columnwidth]{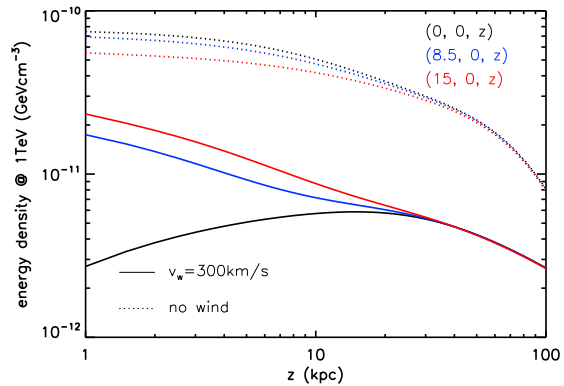}
\caption{Distributions of CR energy density at the present time at 1\,TeV along the axes perpendicular to the Galactic plane, passing through the GC (i.e., point (0, 0, 0), black curves), the earth (i.e., point (8.5kpc, 0, 0), {blue} curves) and the edge of the Galactic plane (i.e., point (15kpc, 0, 0), red curves). The solid and dotted curves respectively represent the cases with the presence of a large-scale radial Galactic wind of $v_w=300\,$km/s and without the presence of the wind. }\label{fig:crden}
\end{figure}

Once we have the CR distribution and gas distribution in the halo, we can calculate the gamma-ray emissivity ($\rm GeV~cm^{-3}s^{-1}$) at $\bm{r}$, which is denoted by
\begin{equation}
J_\gamma(E_\gamma,\bm{r})\equiv \frac{dN_\gamma}{dE_\gamma dt}=cn_{\rm g}(\bm{r})\int_{E_\gamma}^{\infty}\sigma_{pp}N(\bm{r},E)F_\gamma(\frac{E_\gamma}{E},E)\frac{dE}{E}
\end{equation}
following the semi-analytic method developed by \citep{Kelner06}, where $F_\gamma$ is the spectrum of the secondary gamma-ray in a single collision\footnote{see Section IV.A of \citet{Kelner06}, which is based on SIBYLL code \citep{Fletcher94}}.
The total gamma-ray flux average over the solid angle at the Earth can then be given by
\begin{equation}
\begin{split}
\Phi_\gamma(E_\gamma)&=\frac{1}{4\pi(1-\cos70^\circ)}\int\int\int dxdydz \frac{J_\gamma(E_\gamma,\bm{r})}{4\pi (\bm{r}-\bm{r}_E)^2}\\
&\times \left[ 1-\theta\left(3\,{\rm kpc}-|z|\right)\theta\left(15\,{\rm kpc}-\sqrt{x^2+y^2}\right) \right]\\
&\times \theta \left(\sin^{-1} \frac{|z|}{|\bm{r}-\bm{r}_E|}-20^\circ \right)
\end{split}
\end{equation}
where $\theta(x)$ is the Heaviside function. The term in the square bracket is to subtract the emission inside the Galaxy which is regarded as a cylinder with a radius of 15\,kpc and a half height of 3\,kpc above and below the Galactic plane, while the last Heaviside function is to subtract the low-latitude (Galactic latitude $|b|<20^\circ$) emission following the measurement of IGRB by {\it Fermi}-LAT \citep{Fermi15}. That's also the reason that the term $1-\cos 70^\circ$ appears in the denominator of the prefactor of the integration when we average the total flux over the solid angle. Note that given a diffusion coefficient of $10^{29}\rm cm^2/s$, we expect CRs can diffuse to a distance of $2\sqrt{D\Delta t}\simeq 100\,$kpc with a propagation time of $\Delta t=12\,$Gyr, so we only sum up the emission out to $|\bm{r}|=100$\,kpc. Although CRs can travel to a much larger distance in the presence of a Galactic wind, the contribution from larger distances {is} subdominant since both the CR density and the gas density are very low there.

In the calculation of gamma-ray flux, we do not consider the contribution of secondary electrons produced in the $pp$ collision via the decay of charged pions which can emit gamma-rays by inverse Compton scattering off CMB photons. This is because, first, the electron production rate is about half of the gamma-ray production rate in the $pp$ collision of the same parent protons. And, second, to emit TeV photons via inverse Compton scattering off CMB photons, the electron energy needs to be about $10$\,TeV, which are produced by about $200$\,TeV protons in $pp$ collisions, while the required energy of protons that produce TeV photons is about 10\,TeV. Thus, the gamma-ray flux from the secondary electrons is subordinate unless the injection spectrum is hard (i.e., $p<2$). We also neglect the contribution from electrons produced in the electromagnetic cascades initiated by the high-energy photons propagating in the CMB and infrared photon field. {This is because that the mean free path of 20\,TeV photons that  produce 10\,TeV electron/positron pairs is about 100\,Mpc \citep[e.g.][]{Coppi97}, which is much larger than the size of the Galactic halo.}

\begin{figure}[htbp]
\centering
\includegraphics[width=1\columnwidth]{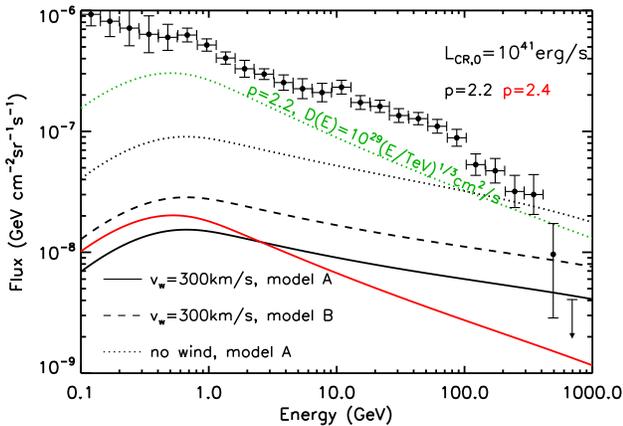}
\caption{Predicted gamma-ray flux from the extended halo of Galaxy. The solid and dashed curves represent the gamma-ray flux with considering a Galactic wind of a constant radial velocity of $300\,\rm km/s$, for the gas density profile model A ($M_g=2.7\times 10^{10}M_\odot$) and model B ($M_g=9.1\times 10^{10}M_\odot$) respectively. The dotted curve represents the case of no wind for the model A. The injection spectral index of CRs is 2.2 for black curves and 2.4 for the red curve. The green dotted curve presents the flux calculated by adopting the diffusion coefficient of the disk for the entire halo. In all the cases shown here, the CR luminosity is $10^{41}\,\rm erg/s$. Filled circles with error bars are IGRB data measured by {\it Fermi}-LAT \citep{Fermi15}. The upper limit at the highest energy bin ($580-820$\,GeV) is shown with the downward arrow.}\label{fig:flux}
\end{figure}

\begin{table*}[htpb]
\centering
\caption{Input parameters in the model.}\label{tab:parameters}
\begin{tabular}{clc}
\hline\hline
Parameters & Descriptions & Values in the fiducial case\\
\hline
$L_{\rm CR,0}$  & Galactic CR luminosity at the present time & $10^{41}$erg/s\\        
$p$ & spectral index of CRs at injection & 2.2\\
$E_{p,\rm max}$ & cutoff energy in the CR injection spectrum & $10^{15}\,$eV\\
$D(E)$ & diffusion coefficient of CR of energy $E$ in the halo & follow Yan \& Lazarian (2008)\\
$v_w$ & velocity of the large scale Galactic wind & $300\,$km/s\\
$n_0$ & halo gas density normalized at 1\,kpc { from} the Galactic center & $0.045\,\rm cm^{-3}$ (Miller \& Bregman 2015)\\
$\beta$ & slope of the density profile & 0.54 (Miller \& Bregman 2015)\\
$S(t)$ & normalized CR injection history with $S(0)=1$ (the present time value) & follow Snaith et al. (2015) (see panel~d of Fig.\ref{fig:setup})\\
\hline
\end{tabular}
\end{table*}

\section{Gamma-ray Flux and the Constraint on the Galactic CR luminosity}
{With the obtained gamma-ray flux, we can put a constraint on Galactic CR luminosity by requiring the gamma-ray flux not to overshoot the observed IGRB flux. On the other hand, the result also depends on parameters other than CR luminosity. Thus, we will explore the influence of these parameters, with a particular focus on the diffusion coefficient.}

{The measured IGRB flux by \emph{Fermi}-LAT is shown with black circles in Fig.~\ref{fig:flux}. The spectrum approximately follows a power law of index of -2.3 and steepens around 100\,GeV. The flux of the highest energy bin at $(580-820)\,$GeV with a center energy of 700\,GeV is consistent with zero and hence an 85\% C.L. upper limit of $4\times 10^{-9}\,\rm GeVcm^{-2}s^{-1}sr^{-1}$ is obtained (corresponding to the upper bound of the $1\sigma$ uncertainty interval) based on the foreground model A in \citet{Fermi15}. Such an upper limit is supposedly to give the most stringent constraint on the model parameters. Note that the foreground model is important to the IGRB flux but it will not effect our result significantly. The upper limit flux will be increased by about a factor of $\sim 1.5$ under other foreground models. }

\subsection{Comparison of expected gamma-ray fluxes in different cases}
Along with the IGRB data, {we compare the predicted gamma-ray flux from the halo in the fiducial case to those with variations in some parameters in Fig.~\ref{fig:flux}.}
The effect of the gas content in the halo can be seen by comparing the black solid curve {(the fiducial case)} and black dashed curve. The gas density model B ($M_g=9.1\times 10^{10}M_\odot$) provides more target atoms for $pp$ collision than the gas density model A ($M_g=2.7\times 10^{10}M_\odot$) by a factor of $\gtrsim 3$. As a result, the gamma-ray flux in the former case is naturally higher than that in the latter case {by about a factor of 2}.

For the fiducial diffusion coefficient and a wind speed of 300\,km/s, convection dominates the CR transportation at $r\gg 4 (D/10^{29}{\rm cm^2s^{-1}})(v_w/300\,{\rm km~s^{-1}})^{-1} $kpc. { Without the Galactic wind, the CR density will be higher and increase the gamma-ray flux (see Fig.~\ref{fig:crden})}. In previous studies \citep{Feldmann13, Kalashev16}, the authors did not consider the Galactic wind. The gamma-ray flux in our ``no wind'' case is comparable to their results although they adopted a higher gas density while we adopted a smaller diffusion coefficient.

The gamma-ray flux is sensitive to the injection {spectrum} of CRs. Comparing the black solid curve to the red solid curve, we can see the flux at the highest energy bin decreases by a factor of 3 when $p$ changes from 2.2 to 2.4. Also, it is obvious that the flux should be proportional to the total CR luminosity at the present time $L_{\rm CR,0}$. 
%

Note that the diffusion coefficient is an important parameter to the result as we mentioned earlier, since the flux is roughly proportional to $1/D$ in the region where diffusion dominates the CR transportation. For reference, we show the gamma-ray flux by applying the diffusion coefficient $D(E)=10^{29}(E/1\,\rm TeV)^{1/3}\rm cm^2/s$ without wind to the entire halo (green dotted curve in Fig.~\ref{fig:flux}). A comparable gamma-ray flux at $\lesssim$TeV is obtained since the diffusion coefficient for $\lesssim 10\,$TeV CRs are similar. {The diffusion coefficient in this case for CRs with $E<10$GeV is one order of magnitude smaller than that of \citet{Yan08}. As a result, the GeV gamma ray emission in this case is higher}. 

\begin{figure}[htbp]
\centering
\includegraphics[width=1\columnwidth]{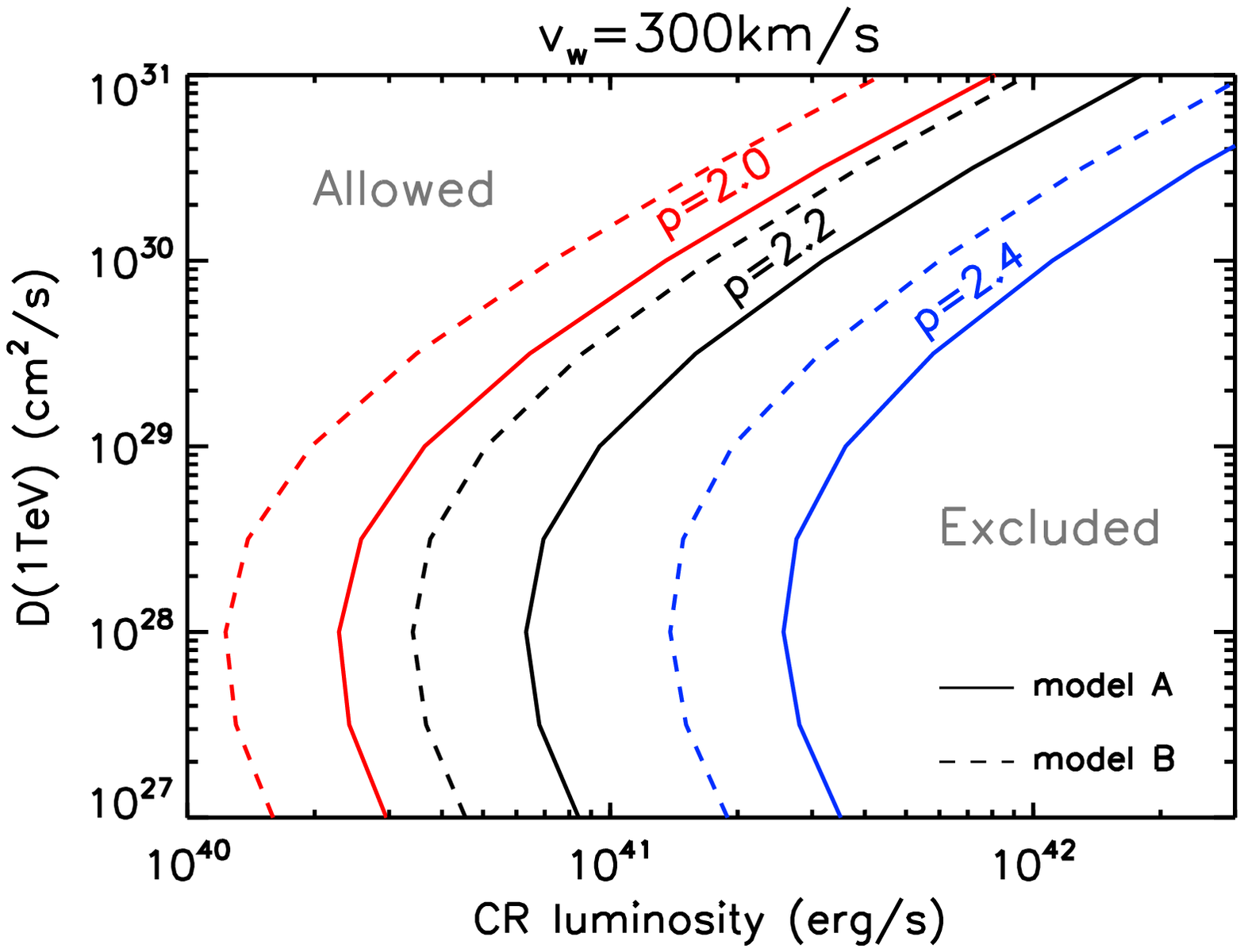}
\includegraphics[width=1\columnwidth]{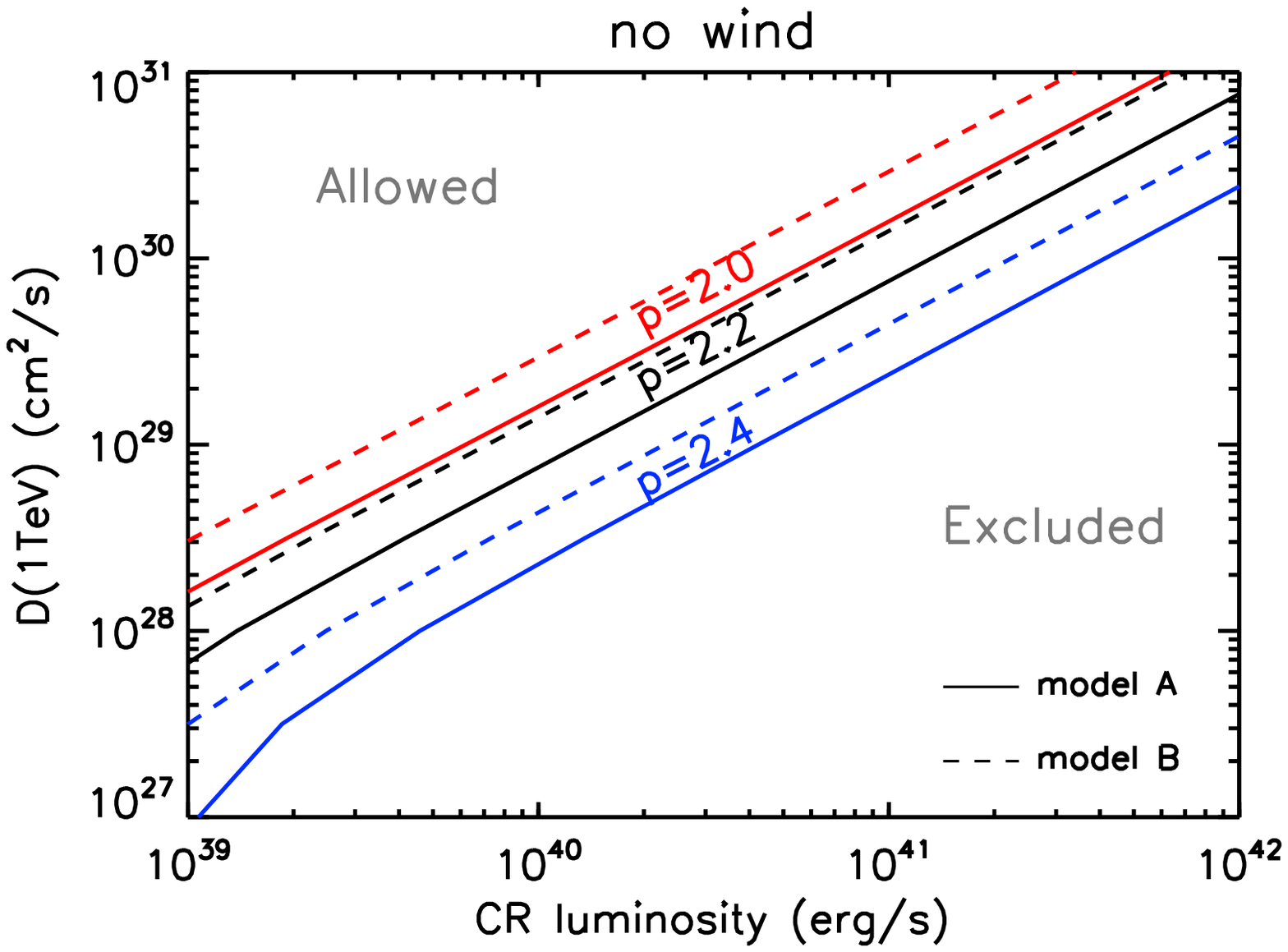}
\caption{The maximum CR luminosity that will not overshoot the IGRB upper limit at $700$\,GeV as a function of the diffusion coefficient in the halo. The diffusion coefficient is assumed to take the form of $D(E)=D(1\,{\rm TeV})(E/1\rm \, TeV)^{1/3}$. Solid curves represent the results with the model A of the halo gas density while dashed curves represent that with the model B. Three different injection spectral indexes are considered (red curves for $p=2.0$, black curves for $p=2.2$ and blue curves for $p=2.4$). {\bf Upper panel}: a radial Galactic wind with a constant speed of $v_w=300\,$km/s is assumed; {\bf Lower panel}: no wind is { present}.}\label{fig:constraint}
\end{figure}

\subsection{Parameter space exploration}
{We note that there are some uncertainties in the model parameters for CR transportation. First,
our assumption of a constant radial Galactic wind may underestimate the gamma-ray flux. 
In fact, whether a large-scale Galactic wind can be launched is still uncertain \citep[e.g.][]{Dubois08}. 
On the other hand, the wind velocity is probably perpendicular to the Galactic plane at $z\lesssim 15$\,kpc rather than radial so that CRs will not suffer adiabatic loss (although the wind may open to a spherical shape at large Galactocentric radius, see \citealt{Zirakashvili96, Recchia16}). Also, the profile of the wind speed is unlikely to be a constant. According to the calculations in previous literature \citep[e.g.][]{Breitschwerdt91, Zirakashvili96, Recchia16}, the wind speed varies with the distance from the Galactic plane and in most region the wind speed is smaller than 300\,km/s, especially near the Galactic plane \citep[see also][]{Taylor17}. {In addition, cosmological simulations suggest a weaker Galactic wind at earlier time \citep{Vogelsberger13, Muratov15}}. 
Apparently, our assumption of the Galactic wind leads to an unrealistically efficient adiabatic loss of CRs at small Galactocentric radius and an overly fast transport of CRs to outer halo by the wind. Nevertheless, the assumption of a constant wind enables a simple analytic solution to the CR transport equation (Eq.~(\ref{eq:crdis})). Thus, instead of employing a more realistic Galactic wind, we propose that the gamma-ray fluxes obtained in the case of no Galactic wind and in the case of a constant wind velocity $v_w=300\,$km/s represent, respectively, an upper bound and a lower bound for the gamma-ray flux in the realistic Galactic wind case.}

Second, the employed benchmark diffusion coefficient \citep{Yan08} is calculated based on the condition of turbulence in the inner halo ($r\lesssim 10\,$kpc, with $n\sim 10^{-3}\rm\,cm^{-3}$, $B\sim \mu$G, $T\sim 10^6\,$K, turbulence injection scale $L\sim 100$pc) and results in an diffusion coefficient of $\gtrsim 10^{29}\, \rm cm^2s^{-1}$ for CRs of $\lesssim 10\,$TeV which is close to the diffusion coefficient in the Galactic disk. This is consistent with { results of} previous studies on the $\lesssim 10\,$kpc radio halo/extended disks of our Galaxy and nearby spiral galaxies \citep[e.g.][]{Orlando13, Mulcahy16, Heesen18}. However, the diffusion coefficient in the extended halo or the outer halo is less known, and it is not necessarily the same as that in the inner halo. Thus, in the { remainder} of this section, we focus on exploring how the diffusion coefficient affect the constraint on CR luminosity.

Since our treatment for CR transportation in this work is only applicable to a spatially independent diffusion coefficient, a homogeneous diffusion coefficient is still adopted for the entire halo. We then assume $D(E)=D(1\,{\rm TeV})(E/1\,\rm TeV)^{1/3}$ with $D(1\,\rm TeV)$ being a free parameter, and calculate gamma-ray flux with different spectral indexes of CRs at injection ($p$), and obtain an upper limit for the CR luminosity at present time by normalizing the predicted gamma-ray flux to the IGRB upper limit at the highest energy bin of $500-820\,$GeV with the central energy $700\,$GeV. The results are shown in Fig.~\ref{fig:constraint}. For each curve, the left side is the allowed parameter space while the right side is the excluded one. Basically, the smaller the diffusion coefficient is, the more stringent the constraint on the CR luminosity will be. The turnovers in the curves when the diffusion coefficient is small are caused by two reasons: first, CRs are trapped in the wind when the diffusion coefficient is small so adiabatic losses of CRs are severe; second, even if there is no wind, since we do not count the gamma-ray produced at small latitude ($|b|<20^\circ$) and inside the Galaxy (which is considered as a cylinder with a radius of 15\,kpc and a half height of 3\,kpc above and below the Galactic plane) into the halo contribution, a non-negligible fraction of CRs still stay in this region and hence the total CR energy budget in the halo is reduced.
Given that the average slope of CR injection spectral index inferred from gamma-ray emissions of local galaxies is suggested to be $p=2.1-2.2$ \citep{Neronov17, Yang16}, it is clear that the current measurement on IGRB already approaches an interesting level to give a nontrivial constraint on CR luminosity for $D(1\,\rm TeV)\leq 10^{30}\,\rm cm^2/s$. From Fig.~\ref{fig:constraint} we see that for $D(1\,\rm TeV)\leq 10^{29}\,\rm cm^2/s$, the CR luminosity can be even restricted to be much smaller than the standard Galactic CR luminosity, i.e. $\sim 10^{41}\,$erg/s, especially in the ``no wind'' case.}

\begin{figure}[htbp]
\centering
\includegraphics[width=1\columnwidth]{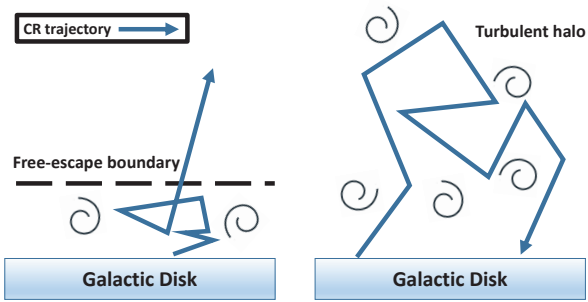}
\caption{Cartoon illustrating the difference of CR propagation with a free-escape boundary (left) and with a turbulent halo surrounding the Galaxy (right). { See Section 4.1 for more discussion.} \label{fig:cartoon}}
\end{figure}

We note that the upper limit for CR luminosity obtained in Fig.~\ref{fig:constraint} may be quite conservative, because the IGRB is expected to be dominated by unresolved extragalactic sources\citep[][see also the analysis in \citealt{Liu16}]{Fermi16, Ajello15, Lisanti16}, such as BL Lacs \citep{DiMauro14b} and radio galaxies \citep{DiMauro14, Hooper16}. The room left for the halo contribution could be just a small fraction of the value of the IGRB upper limit. {For example, if 90\% of the IGRB upper limit at 700\,GeV turns out to be contributed by unresolved sources, the allowed Galactic CR luminosity will be 10 times smaller than the obtained value in Fig.~\ref{fig:constraint} for the same diffusion coefficient}. We expect the next generation very-high-energy gamma-ray detectors, such as CTA, {are} able to resolve more extragalactic TeV gamma-ray sources from the IGRB { providing/allowing a more accurate limit for the halo contribution}. If so, the constraint on the Galactic CR injection would become stringent and provide useful clues to the origin of Galactic CRs. For example, provided a smaller CR luminosity, the required acceleration efficiency of supernova remnants could be lower, and other potential CR accelerators such as OB associations \citep{Cesarsky83, Bykov01, Aharonian18}, pulsars \citep{Bednarek04}, and Galactic center \citep{HESS_GC16} {become} possible to account for the majority of Galactic CRs in terms of energy budget.

%


\begin{figure}[htbp]
\centering
\includegraphics[width=1\columnwidth]{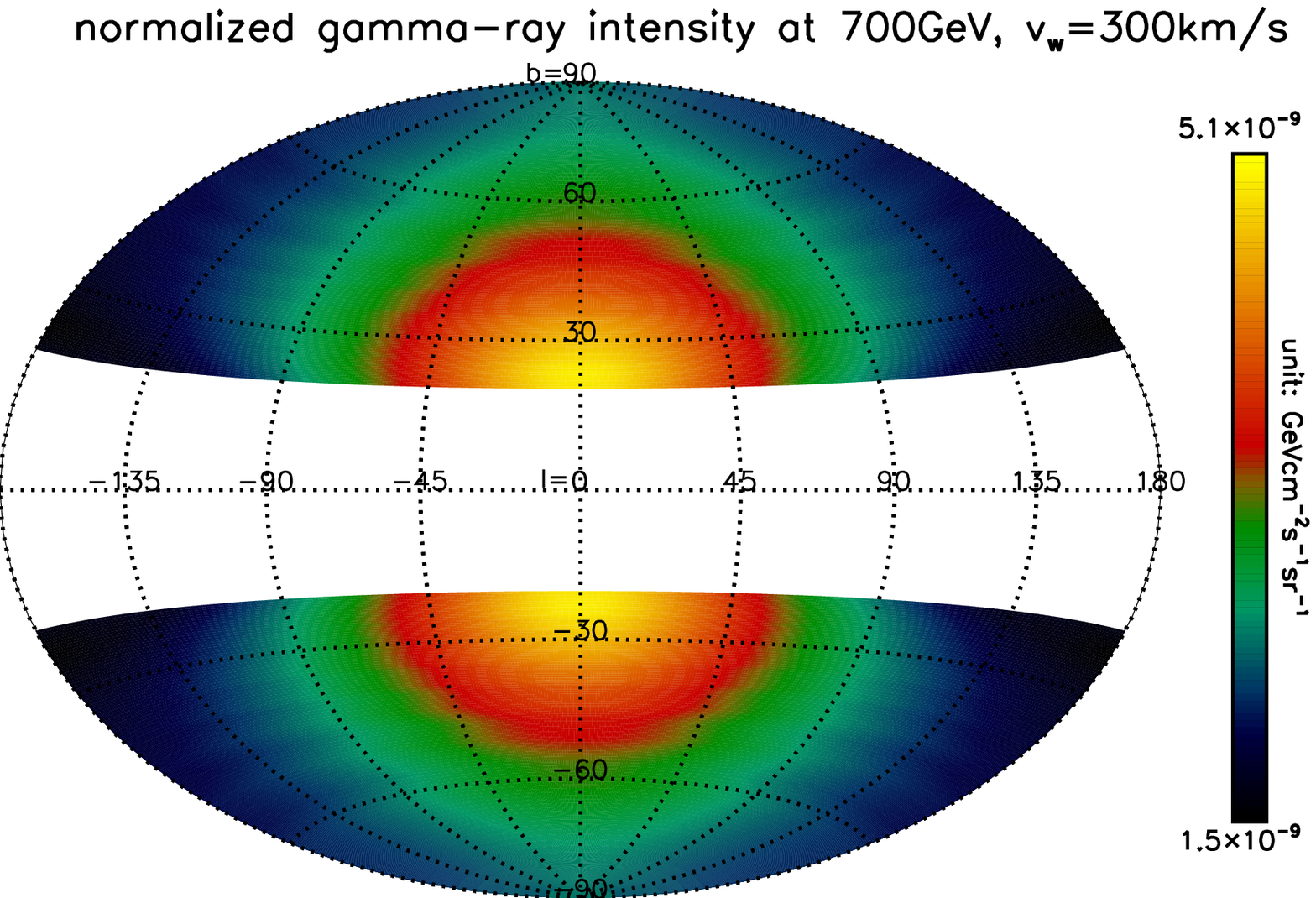}
\includegraphics[width=1\columnwidth]{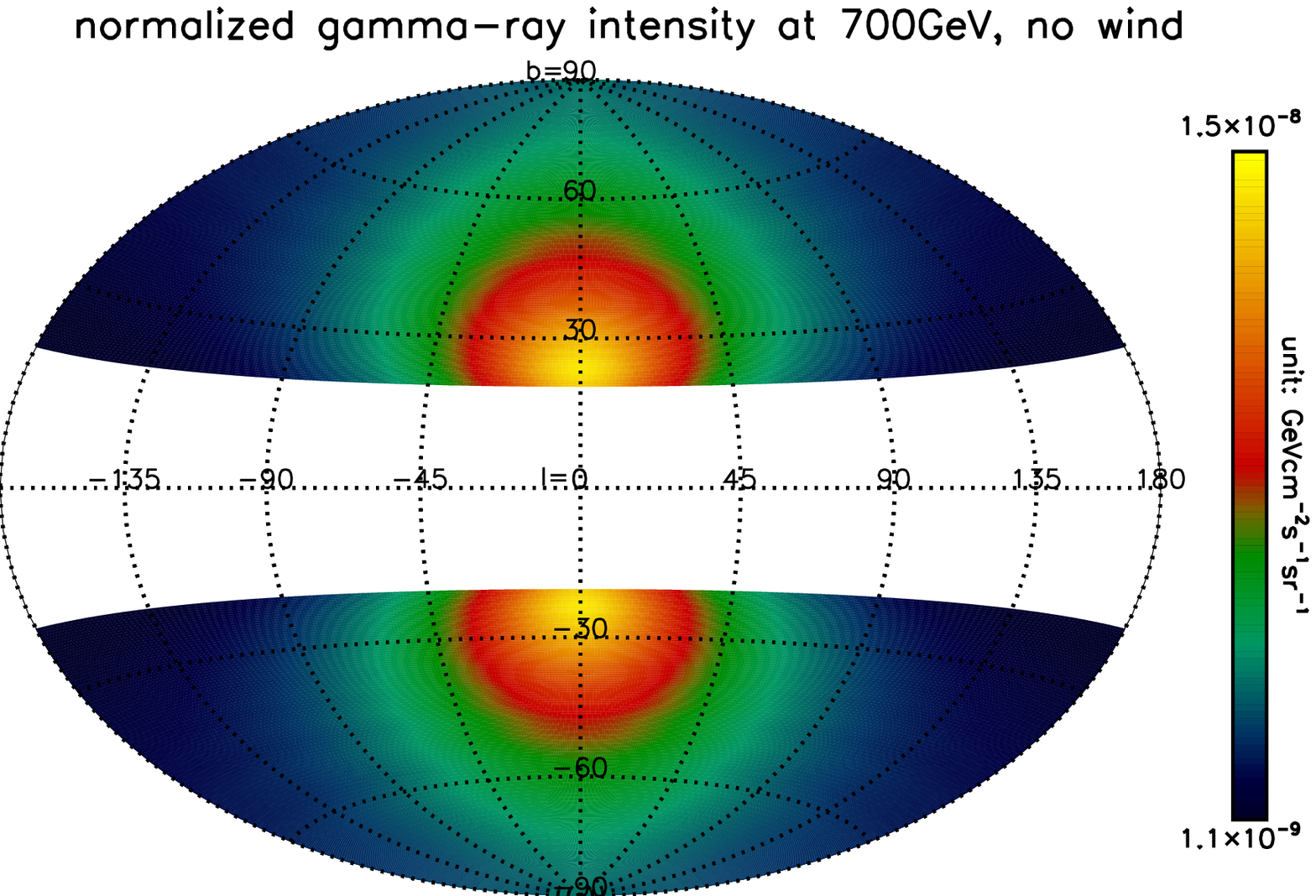}
\caption{Sky map in Galactic coordinates of high-latitude ($|b|>20^\circ$) gamma-ray intensity at 700\,GeV produced in the halo. The upper panel shows the case with a large-scale Galactic wind of $v_w=300\,$km/s while no wind appears in the lower panel. The injection spectral index of CRs is $p=2.2$ in both panels. The {all-sky average gamma-ray intensity} is normalized to the upper limit of highest-energy bin of the IGRB measured by {\it Fermi}-LAT. The color scale is linear. {See Section 4.2 for details.}}\label{fig:image}
\end{figure}

\section{Discussion}
\subsection{Alternative ways to reduce the sub-TeV gamma-ray flux produced in the halo}
{A free-escape boundary of CRs is considered in many previous literature that study the CR transport in the Galaxy. Such a boundary is usually assumed to locate at several kpc above and below the Galactic plane. Once CRs cross the boundary, they will leave the Galaxy and never return, and the CR density is also imposed to be zero at the free-escape boundary {\citep[in many numerical studies such as][]{Strong98, Evoli08, Kissmann14}}. The estimated CR luminosity based on this model is usually in the range of $(0.3-3)\times 10^{41}\,$erg/s. In the previous section, we have shown that in the presence of a turbulent halo, the sub-TeV IGRB upper limit may imply a smaller Galactic CR luminosity than the conventional value based on the assumption of a free-escape boundary. This is because that the physical picture of the CR transportation in these two scenarios is different (see Fig.~\ref{fig:cartoon}). CRs that have propagated into the halo that is far away from the Galactic plane may still have a chance to return back if the halo is turbulent, especially if there is no large-scale convective Galactic wind or the wind is weak. As a result, the required CR luminosity to maintain the locally observed CR energy density is also smaller than that in the case with a free-escape boundary.} 

{There are also other mechanisms} to reduce the theoretical gamma-ray flux at sub-TeV energy, without invoking a small Galactic CR luminosity. An {apparent} solution is to employ a large diffusion coefficient in the extended halo and a soft injection spectrum (e.g.,$p=2.4$). As can be seen in Fig.~\ref{fig:constraint}, such a combination relax the constraint on the CR luminosity significantly.  
On the other hand, IGRB flux at sub-Tev energy actually only constrains the luminosity of $\sim 1-10$\,TeV CRs, which do not necessarily originate from the same sources of GeV CRs where most of the CR energy resides. There must be less TeV CRs sources than GeV CR sources since the requirement for accelerating higher energy CRs is more demanding. It is possible that the measured TeV CRs are subject to a few nearby sources rather than the sources in the entire Galaxy, such that the total TeV CR luminosity is smaller than the currently inferred value which is based on the assumption that TeV CR flux are the same in the entire Galactic disk as that measured at the Earth. Consequently, the sub-TeV gamma rays produced in the halo can be reduced effectively. This scenario is implied by the recent discovery of the spectral hardening of Galactic CRs at $\sim 200$GV \citep[e.g.][]{Panov09, Yoon11, Adriani11, Aguilar15}, since CR spectrum from closer and younger sources tend to be harder since the spectrum is less effected by the energy dependent diffusion. The anisotropy study on TeV CRs may also support such a scenario \citep{Ahlers16}.

\subsection{Anisotropy of Gamma-ray Intensity}
As pointed out in previous studies \citep{Feldmann13, Kalashev16}, we can expect large-scale anisotropy in the intensity of gamma rays produced in the halo at different Galactic longitude $l$ and Galactic latitude $b$, unlike that originated from extragalactic sources which is supposed to be roughly isotropic. Particularly, due to the offset of the Earth from the GC, the gamma-ray intensity should be enhanced towards the GC direction ($l=0^\circ$) and is decreased towards the anti-GC direction ($l=180^\circ$). We here present the intensity map of 700\,GeV gamma-rays produced in the halo by integrating the gamma-ray flux in the light of sight for each $l$ and $b$. As exhibited in Fig.~\ref{fig:image}, the upper panel shows the case with a wind speed of $300\,$km/s while the lower panels shows the ``no wind'' case. In both panels, the injection indexes of CRs $p=2.2$ are adopted, and {the all-sky averaged  gamma-ray intensity} is normalized to the upper limit of the IGRB at the highest-energy bin measured by {\it Fermi}-LAT. We can see the gamma-ray intensity varies with $l$ and $b$. The presence of a large-scale Galactic wind with spherical symmetry tends to reduce the anisotropy, since the convection by the wind is isotropic and more CRs are transported to larger distance in the anti-GC direction. Remarkably, due to a higher CR density and gas density in the inner halo, the gamma-ray emissivity is relatively high and we can see a bubble-like structure above and below the GC in {both cases of $v_w=300\,\rm km/s$ and ``no wind''}. {We have also calculated the gamma-ray intensity map with employing, instead of a radial wind, a wind perpendicular to the Galactic plane for the inner 15\,kpc region and a similar bubble-like structure appears all the same.} 


We note that whether such a structure is related to the Fermi bubble \citep{Su10, Yang14}, which is mainly detected at $10-100\,$GeV with a much higher intensity, remains to be studied in detail. Given the constraint of the IGRB upper limit at 700\,GeV, the 10\,GeV intensity of the bubble-like region in Fig.~\ref{fig:image} is about one order of magnitude smaller than that of the Fermi bubble.  
Nevertheless, we note that the real situation is much more complex. For example, the intensity of the bubble-like region can be enhanced if we consider an additional CR injection from the GC region due to its past activity \citep{Guo12, Barkov14, Crocker15}, which is inferred from X-ray observation\citep{Revnivtsev04, ZhangS15} and the discovery of PeV proton accelerator at GC \citep{HESS_GC16, Liu16, Fujita17}. Besides, the intensity and the morphology of the bubble-like structure can be influenced by the diffusion coefficient or possible outflow in the bubble region, which are not necessarily the same with those in the rest part of the halo. Also, 10\,GeV gamma rays are mainly produced by 100\,GeV protons, at which energy the CR self-regulation via scattering off self-excited Alfv{\'e}n waves most likely have an influence on the CR transport and result in a smaller diffusion coefficient than the benchmark one. A detailed study on the Fermi bubble is, however, beyond the scope of this work. 

\begin{figure}[htbp]
\centering
\includegraphics[width=1\columnwidth]{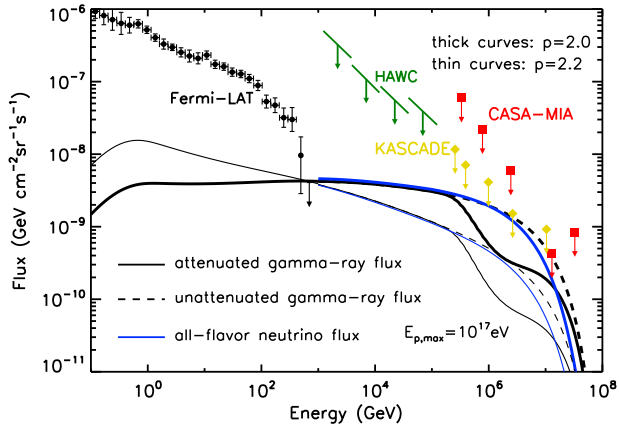}
\caption{Gamma-ray flux (black curves) and neutrino flux (blue curve) produced in the Galactic halo, assuming maximum CR energy at injection to be $E_{p,\rm max}=10^{17}\,$eV. The gamma-ray fluxes are normalized to the upper limit of the IGRB measured by {\it Fermi}-LAT at 700\,GeV. Thick curves represent the results with $p=2.0$ while thin curves represent the results with $p=2.2$. Black solid curves are the gamma-ray flux after considering the absorption by the CMB and the black dashed ones are those before the absorption. Black filled circles with error bars represent the IGRB measured by {\it Fermi}-LAT \citep{Fermi15},  green slashes show the upper limit of diffuse gamma-ray flux from the northern Fermi bubble region measured by HAWC \citep{HAWC17}, yellow squares  show the upper limit of the isotropic diffuse gamma-ray flux measured by KASCADE \citep{KASCADE03}, and red dimonds show the upper limit of the isotropic diffuse gamma-ray flux measured by CASA-MIA \citep{CASAMIA97}. }\label{fig:flux_HE}
\end{figure}
\subsection{Gamma-ray and Neutrino Production at TeV-PeV}
In the calculation above, we assume a maximum CR proton energy $E_{p, \rm max}=10^{15}\,$eV at injection. The maximum CR energy attainable in the Galactic sources could be much higher so that we may expect the gamma-ray spectrum to extend well beyond TeV energy. High-energy neutrinos can also be produced in the $pp$ collision. In Fig.~\ref{fig:flux_HE}, we show the expected gamma-ray flux and neutrino flux produced in the halo by assuming $E_{p,\rm max}=10^{17}\,$eV, {which is the suggested maximum proton energy achievable in supernova remnants\citep{Ptuskin03}}, and adopting gas density model A and the fiducial diffusion coefficient.The attenuation of gamma-ray by the CMB is considered, while cosmic infrared background and the infrared radiation from our Galaxy are not important to the attenuation and hence neglected. The gamma-ray flux is normalized to the IGRB upper limit at 700\,GeV {to show the maximum neutrino flux produced in the halo, so it makes no difference on the resulting gamma-ray/neutrino flux whether we consider a Galactic wind or not.} Compared to the constraints from observations of HAWC, CASA-MIA and KASCADE above TeV, the constraint from {\it Fermi}-LAT is the most stringent as long as $p\geq 2$. {Note that, however, if the maximum proton energy is larger than $10^{18}\,$eV, the gamma-ray flux will extend beyond $10^{17}\,$eV, at which energy the attenuation due to CMB becomes weak since the attenuation length of $10^{17}\,$eV photon is about $\sim 100\,$kpc which is comparable to the size of the halo. In this case, if the injection spectrum is hard, e.g., $p=2$, the CASA-MIA data may be more constraining than the {\it Fermi}-LAT data.}

Given the constraint by $\lesssim$TeV IGRB, the all-flavor flux of high-energy neutrinos produced in the halo can at most reach the same level of the upper limit of IGRB at $700\,$GeV. Assuming a flavor ratio of $1:1:1$ among the three flavors after oscillation, we find the obtained per-flavor neutrino flux can contribute at most about $3\times 10^{-9}\rm \,GeVcm^{-2}s^{-1}sr^{-1}$ or $(10-30)\%$ of the measured flux by IceCube at 100\,TeV, i.e., $\simeq (1-3)\times 10^{-8}\,\rm GeV~cm^2s^{-1}sr^{-1}$ (depending on whether we compare with the high energy starting events data or the through-going muon data \citealt{IC15, IC17_ICRC, Palladino18}), unless the Galactic CRs are injected with a harder spectrum ($p<2$). This result is consistent with the conclusions in previous literature \citep[e.g.][]{Ahlers14, Murase16}. Note that the constraint on the neutrino flux put by the $\lesssim$TeV IGRB is weaker for more distant sources \citep{Chang16}, since the attenuation of $\lesssim$TeV gamma-rays by the cosmic infrared background will be more important.

\subsection{Baryon budget of hot CGM of our Galaxy}
The cosmic mean baryon fraction is found to be $\approx 16.5\%$ from the CMB observation with high accuracy \citep{Hinshaw13}. However, the total observed baryonic mass within the virial radius the Galaxy is much smaller than the inferred mass from the cosmic mean baryon fraction. This has motivated a great deal of efforts of uncovering the baryons that are ``missing'' from the Galaxy \citep[for review see][]{Bregman07, Tumlinson17}. There have been predictions that the missing baryons are residing in a hot state in the Galactic halo \citep{White78,White91}, which is known as the CGM. While \citet{Bregman18} suggests the hot gas does not account for the missing baryons by a factor of $3-10$, the total mass of the hot gas in the halo, however, is actually far from settled since the conclusion is suffered from the uncertainties of metallicity \citep{Faerman17, Qu18}, radiative transfer effects \citep{LiYY17}, halo rotation \citep{HK16}, {the size of the halo}, and the possible flattening of the gas distribution with respect to the $\beta$ model at the larger radius \citep{Gupta12, Faerman17}. For example, with flatter density profile at larger radii, \citet{Faerman17} suggest the mass of warm/hot gas is as high as $1.2\times10^{11}M_\odot$ which may account for the missing baryons in the Galaxy, although \citet{Bregman18} argue that the flattening of $\beta<0.4$ is not consistent with the \ion{O}{7} column observations toward LMC and SMC.

\begin{figure}[htbp]
\centering
\includegraphics[width=1\columnwidth]{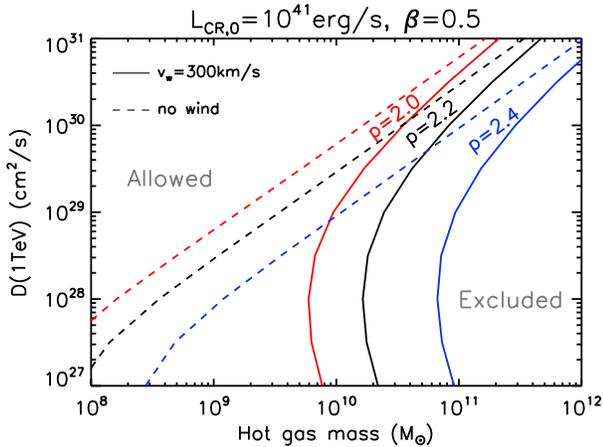}
\caption{The maximum hot gas mass in the halo that will not overproduce the IGRB flux at 700\,GeV. Solid curves show the results with a wind speed $v_w=300$km/s while dashed curves show the results without wind. Red, black, blue curves represent the results with $p=2.0,2.2,2.4$ respectively. In the figure, the present time CR luminosity is fixed at $10^{41}$erg/s and $\beta=0.5$ is adopted for the gas density profile in the halo.}\label{fig:constraint_mass}
\end{figure}

The gas mass and their distribution in the halo is apparently important to the production of $\gamma$-ray background, therefore our results may in turn provide an independent constraint on the baryonic component of Galactic gaseous halo, if the properties of Galactic CRs can be fixed.
For example, {in the fiducial case},  we find that the mass of the hot gas halo cannot exceed $\sim2.6\times10^{10}M_\odot$, which is only {$\sim 11\%$ of the missing baryon mass in the Galaxy given a total mass of $\sim1.4\times10^{12}M_\odot$ \citep{Watkins18} for the Galaxy}, in order not to overshoot the IGRB upper limit at $\sim700\,$GeV. {We note that the result is subject to the uncertainty of various model parameters. We then explore the influences of the diffusion coefficient, the wind speed, and the injection CR spectral index on the result in Fig.~\ref{fig:constraint_mass}. The CR luminosity in the calculation is fixed at $10^{41}\,$erg/s and we do not explore the dependence on it since upper limit of the gas mass is simply proportional to $L_{\rm CR,0}^{-1}$. In the plot, we fix $\beta$ at 0.5 and adjust the amplitude in the gas density profile ($n_0$) to make the predicted gamma-ray flux at 700\,GeV equal to the measured upper limit. The maximum gas mass can then be calculated by integrating gas density profile over the entire volume within the virial radius. Similar to Fig.~\ref{fig:constraint}, the region to the left region of a curve is the allowed parameter space in the corresponding condition of the curve.} Considering that the Galactic halo only contributes a fraction of the total IGRB, the constraint to the halo gas mass will be even stronger if a better estimation of the extra-galactic contribution and/or a more accurate upper limit of the IGRB at $\sim700$~GeV are obtained. {If the total gas mass in the halo is fixed, a smaller $\beta$ leads to a lower gas density than the case of a large $\beta$ in the inner halo where CR density is relatively high and the distance to the Earth is small. As a result, the expected gamma-ray flux from the halo will be smaller, allowing a larger amount of gas in the halo. Hence, all the curves in Fig.~\ref{fig:constraint_mass} will shift to the right. }

\section{Conclusion}
To summarize, we calculated the distribution of Galactic CRs in the extended halo after they leave the Galaxy, and the gamma-ray production by these CRs via the $pp$ collision with the gas in the halo. Given a total gas mass of $10^{10}-10^{11}\,M_\odot$ in the halo, we found that the current measurement of the IGRB at $\lesssim$TeV by {\it Fermi}-LAT already approaches a level that may provide an independent non-trivial constraint on the luminosity of Galactic CRs.  {In our fiducial case, the present time Galactic CR luminosity is required to be smaller than $10^{41}\,$erg/s, in order not to overshoot the measured IGRB upper limit with the predicted gamma-ray flux produced in the halo. Since the IGRB is expected to be dominated by extragalactic sources, future observations on the extragalactic gamma-ray sky may further shrink the room left for the Galactic halo contribution and provide {an even stronger constraint on the Galactic CR luminosity and their origin}. A small Galactic CR luminosity is consistent with the presence of an extended turbulent halo surrounding the Galaxy, since in this scenario CRs that leave the Galaxy will stand a chance to return. 
{Our constraint on CR luminosity is} influenced by various model parameters. We found that a large diffusion coefficient in the extended halo (i.e., $D>10^{29}\rm \, cm^2/s$ at $E=1\,\rm TeV$) or a soft injection spectrum of CR (i.e., $p>2.2$) can reduce the predicted gamma-ray flux, relaxing the constraint on the CR luminosity. An alternative possible scenario which can reduce the predicted sub-TeV gamma-ray flux is the presence of a local overdensity of TeV CR source.}

{A few more implications of our results may be worth noting. First,} the interaction between CRs injected from Galactic plane and gas in the halo can produce a bubble-like structure in gamma-ray above and below the Galactic plane. Given the constraint from the IGRB upper limit at $\lesssim \rm TeV$, the intensity of the bubble-like structure is much smaller than that of the Fermi bubble. Additional CR injection from the GC and a distinct environment in the region (e.g., a smaller diffusion coefficient {or a higher gas density}) from the rest part of the halo may help to explain the Fermi bubble in this scenario. {Second,} due to the constraints by the IGRB upper limit, high-energy neutrinos produced in the halo can reach at most $(10-30)$\% of the measured flux by IceCube above 100\,TeV. {Last,} if the properties of the Galactic CRs can be fixed, the IGRB upper limit may also be useful to constrain the baryon budget of the hot gas in the halo. {For example, our fiducial model constrains the hot gas mass in the halo to be lower than $2.6\times 10^{10}\,M_\odot$. However, this value is subject to the adopted CR spectral index at injection, diffusion coefficient {as well as the gas profile in the halo}. For $p=2.4$, $D(1\,\rm TeV)=10^{29}\rm\,cm^2/s$ {and $\beta=0.5$}, the maximal mass is $10^{11}M_\odot$. }

\begin{acknowledgements}
We thank Felix Aharonian and the anonymous referee for invaluable comments, and thank Martin Pohl, Christoph Pfrommer and Susumu Inoue for helpful discussions. This work is partially supported by the National Key R \& D
program of China under the grant 2018YFA0404203 and the NSFC grant 11625312 and 11851304.
\end{acknowledgements}

\bibliography{ms.bib}
\end{document}